\documentclass{PoS}

\usepackage{lineno}
\usepackage{float}
\title{Charm baryon production in pp, p-Pb and Pb-Pb collisions with ALICE at the LHC}

\newcommand{\lc}{$\Lambda_{\rm c}~$}
\newcommand{\lcp}{$\Lambda_{\rm c}^{+}~$}

\newcommand{\CsiC}{$\Xi_{\rm c}^{0}~$}
\newcommand{\dzero}{$\rm D^{0}~$}
\newcommand{\pt}{$p_{\rm T}$}
\newcommand{\rs}{$\sqrt{\rm s}$}
\newcommand{\rsnn}{$\sqrt{\rm s_{NN}}$}
	
\ShortTitle{}

\author{\speaker{Elisa Meninno on behalf of the ALICE Collaboration}\\
        INFN and University of Salerno, Salerno, Italy\\
        Stefan Meyer Institute for Subatomic Physics (SMI), Vienna, Austria\\
        E-mail: \email{elisa.meninno@cern.ch}}

\abstract{
	In this contribution, the latest ALICE results on charmed baryon production are presented. In particular the measurements of \lcp baryon production in pp collisions at ${\sqrt{\rm s} \,= 5.02 \rm \,TeV}$ and in p-Pb collisions at \rsnn \,= 5.02 TeV are shown and compared with previous measurements performed using pp and p-Pb data samples collected during \mbox{LHC Run 1}. At mid-rapidity charmed baryons are reconstructed via their hadronic decay and semileptonic decay channels, exploiting the high precision tracking, excellent vertexing and particle identification capabilities offered by the ALICE detector. The first measurement of \lcp-baryon production in Pb-Pb collisions at \rsnn \,= 5.02 TeV is also shown. Results are compared with theoretical expectations.}

\FullConference{International Conference on Hard and Electromagnetic Probes of High-Energy Nuclear Collisions\\
		30 September - 5 October 2018\\
		Aix-Les-Bains, Savoie, France}

\begin{document}
\section{Introduction}
The study of charm production is a powerful tool to investigate the Quark-Gluon-Plasma (QGP), the deconfined state of matter created under extreme energy densities in heavy-ion collisions. Heavy quarks are produced in hard parton scattering processes occurring in the early stages of the collision, and as they traverse the QCD medium, they interact with its constituents and thus experience the whole evolution of the medium. In particular, the charmed baryon-to-meson ratio in heavy-ion collisions is sensitive to the charm hadronisation mechanisms in the QGP. Indeed it is expected that a significant fraction of low and intermediate momentum charm and beauty quarks hadronise via coalescence or recombination with the other light quarks in the medium. This produces an enhancement of the \lcp/\dzero ratio with respect to pp collisions. Moreover, a further enhancement is expected if the existence of light di-quark bound states occurs in the QGP~\cite{Lee:2007wr}. 
The interpretation of the results obtained in heavy-ion collisions requires detailed studies also in smaller systems:
pp collisions provide the necessary reference for measurements and allow to test pQCD predictions and hadronisation models at the LHC energies.
Measurements in p-Pb collisions are traditionally used to investigate cold-nuclear matter effects that are related to the presence of nuclei in the colliding system and that could mimic final-state medium-related effects~\cite{Arneodo:1992wf}.

\section{Charm measurements with ALICE}
ALICE is a heavy-ion dedicated experiment at CERN, aimed to study the properties of the QGP. The ALICE detector is characterized by unique capabilities for Particle Identification (PID) and low-momenta tracking~\cite{Aamodt:2008zz}. It consists of a central barrel at mid-rapidity, a muon spectrometer at forward rapidity and a set of detectors for triggering and event characterization. Charmed hadrons are studied in ALICE using both hadronic decay and semileptonic decay channels.\\
%\textbf{Semi-leptonic decay reconstruction}:\\
\CsiC \,and \lcp production cross sections have been measured in pp collisions via their semileptonic decay channels, \CsiC $\rightarrow$ $\rm e^{+} \Xi^{-} \nu_{e}$ \cite{Acharya:2017kfy} and \lcp $\rightarrow$ $\rm e^{+} \Lambda$~\cite{Acharya:2017lwf} respectively. Electrons are identified by combining the information about the energy loss in TPC and the time of  flight in TOF. After the $\Lambda$ ($\Xi$) candidates are reconstructed, $e\Lambda$ ($e\Xi$) pairs are built, applying selections on opening angle and invariant mass. Due to the missing momentum carried by the undetected neutrino, the invariant mass distributions do not show a peak at the \lcp (\CsiC) mass. In this case the \lcp (\CsiC) distributions are obtained subtracting Wrong Sign (WS) pairs from the Right Sign (RS) yields. An unfolding technique is then used to convert the $\rm e \Lambda$ ($ \rm e \Xi$) \pt-spectrum in the desired \lcp (\CsiC) \pt-spectrum. \\
%\textbf{Hadronic decay reconstruction}:\\
\lcp are also reconstructed using their hadronic decay channels \lcp $\rightarrow$ $\rm pK^{-}\pi^{+}$ and \lcp $\rightarrow$ $\rm pK^{0}_{S}$~\cite{Acharya:2017lwf}. \lcp $\rightarrow$ $\rm pK^{-}\pi^{+}$ candidates are built from triplets of selected tracks.
For the \lcp $\rightarrow$ $\rm pK^{0}_{S}$, $\rm K^{0}_{S}$ candidates are reconstructed from pairs of opposite-sign tracks, forming a secondary vertex displaced from the interaction vertex. The \lcp candidates are built combining $\rm K^{0}_{S}$ and $\rm p$ candidates.
For both decay channels, the background is reduced with cuts on kinematical and geometrical variables, tuned for each \pt \,interval.
The identification of the proton for the \lcp $\rightarrow$ $\rm pK^{0}_{S}$ is based on the specific energy loss and time of flight from TPC and TOF detectors, respectively.
For the \lcp $\rightarrow$ $\rm pK^{-}\pi^{+}$ analysis, affected by a larger combinatorial background, a bayesian PID, using TPC and TOF information is used.
In the case of \lcp $\rightarrow$ $\rm pK^{0}_{S}$, the analysis was performed also using a multivariate approach based on Boosted Decision Trees (BDT)~\cite{TMVA2007}.         

\section{Results}
\begin{figure}
	%\begin{center}
	\includegraphics[width=0.5\linewidth]{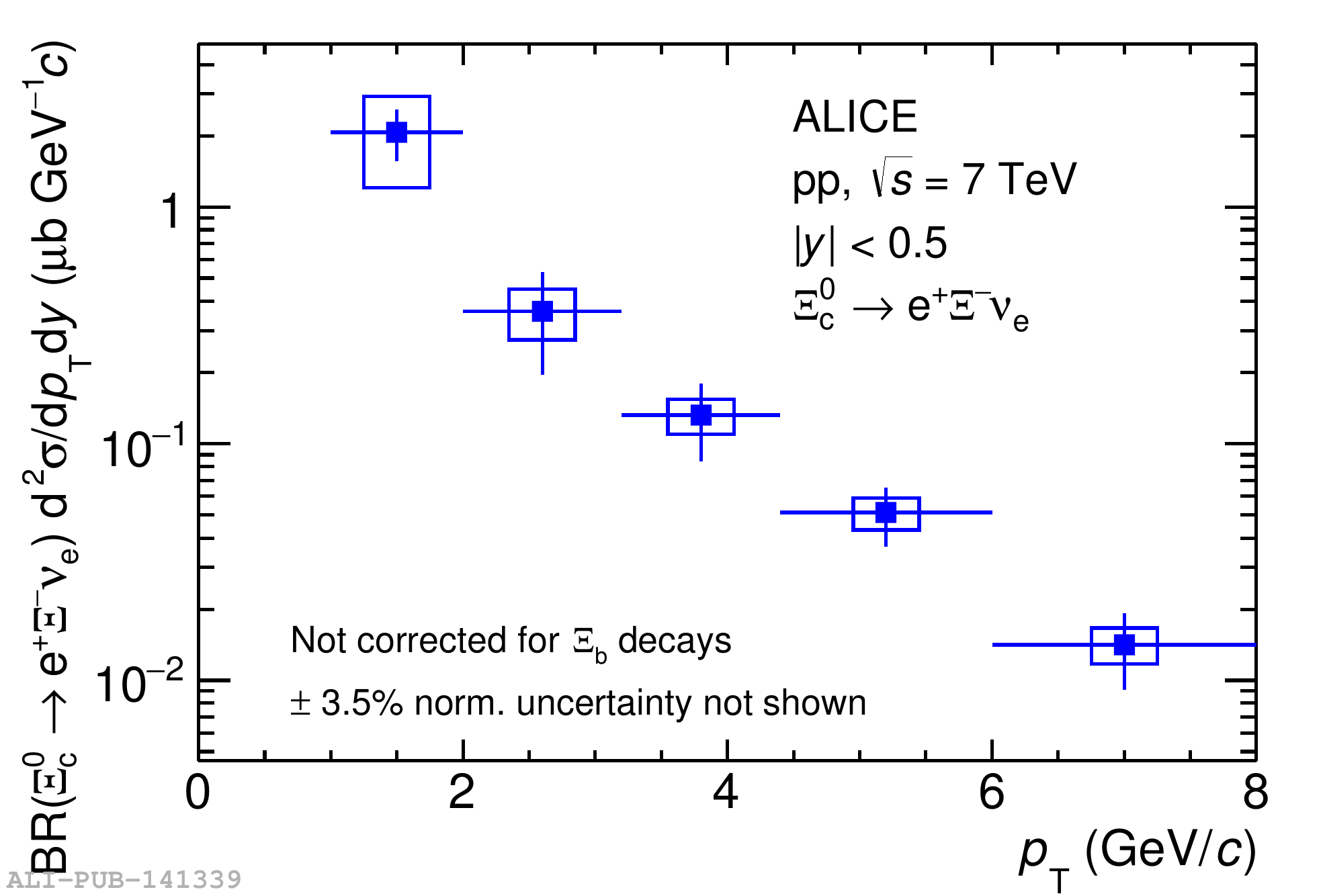}
	\includegraphics[width=0.5\linewidth]{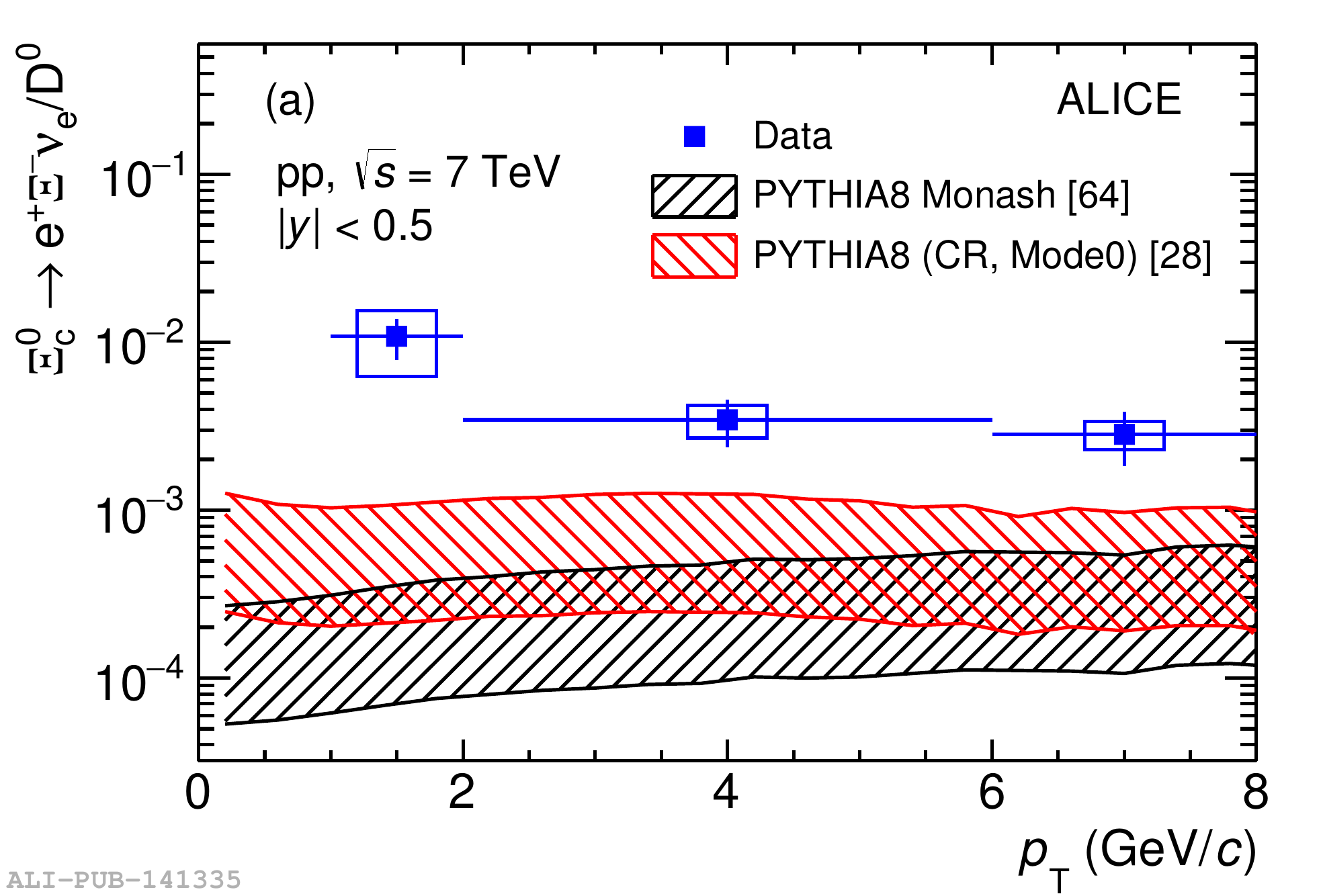}
	\caption{Left: \CsiC-baryon \pt-differential cross section in pp collisions at \rs = 7 TeV, multiplied by the branching ratio into $e^{+}\Xi^{-}\nu_{e}$. Righ: \CsiC/\dzero ratio as a function of \pt, compared with predictions.}
	\label{fig:csi}
	%\end{center}
\end{figure}

The inclusive \CsiC-baryon \pt-differential cross section measured at mid-rapidity in ${1 < p_{\rm T} < 8}$ GeV/$c$, multiplied by the branching ratio into $e^{+}\Xi^{-}\nu_{e}$ is shown in the left panel of Fig.~\ref{fig:csi}. In the right panel, the \CsiC / \dzero ratio is shown and compared with predictions obtained using PYTHIA event generator with different tunes of hadronisation. The CR Mode tune~\cite{Christiansen:2015yqa}, which includes string formation beyond the leading-colour approximation, provides a better description of the data, although the measured baryon/meson ratio is still higher than the expectations.

\begin{figure}
	\begin{center}
	\includegraphics[width=0.5\linewidth]{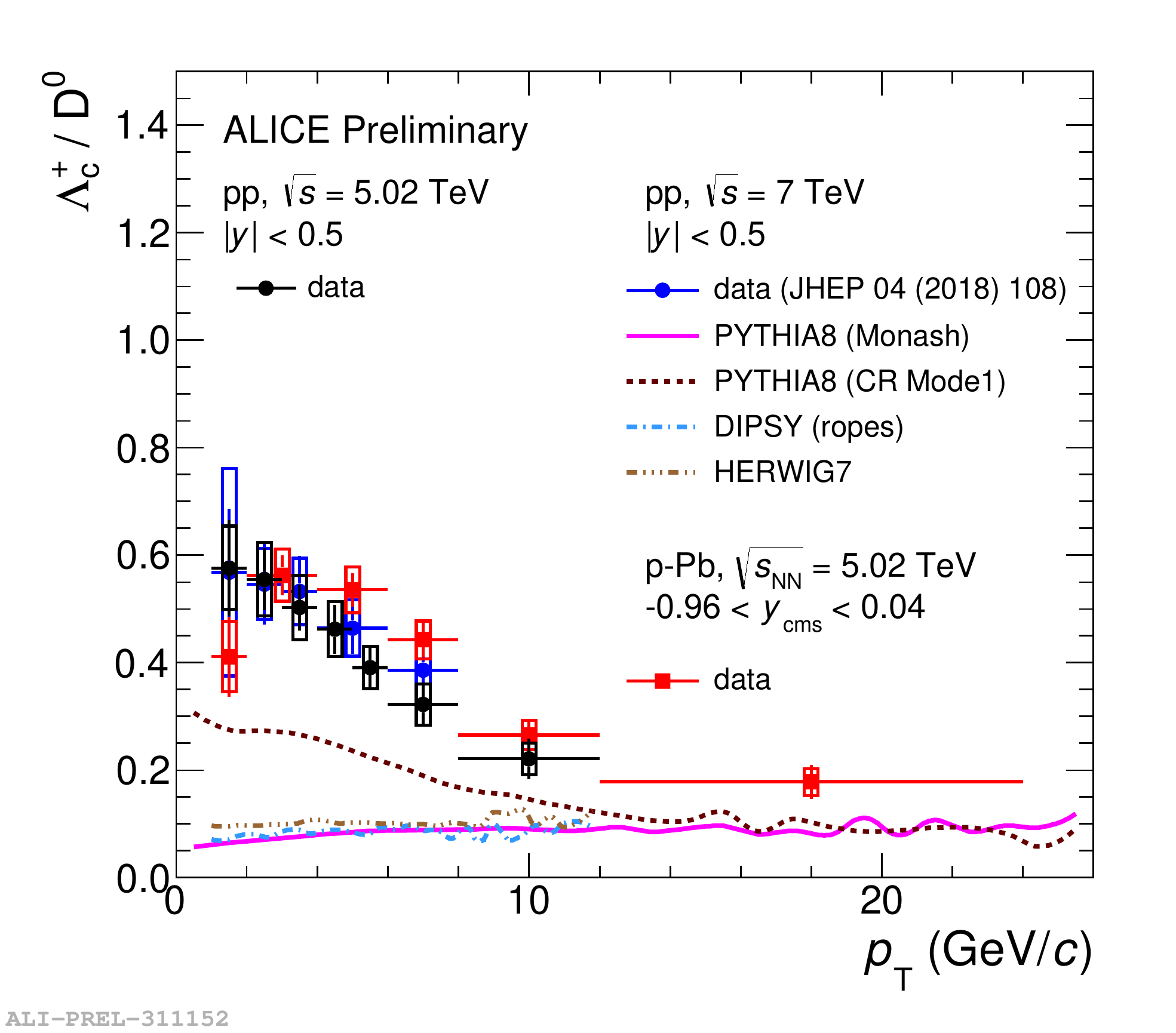}
	\includegraphics[width=0.9\linewidth]{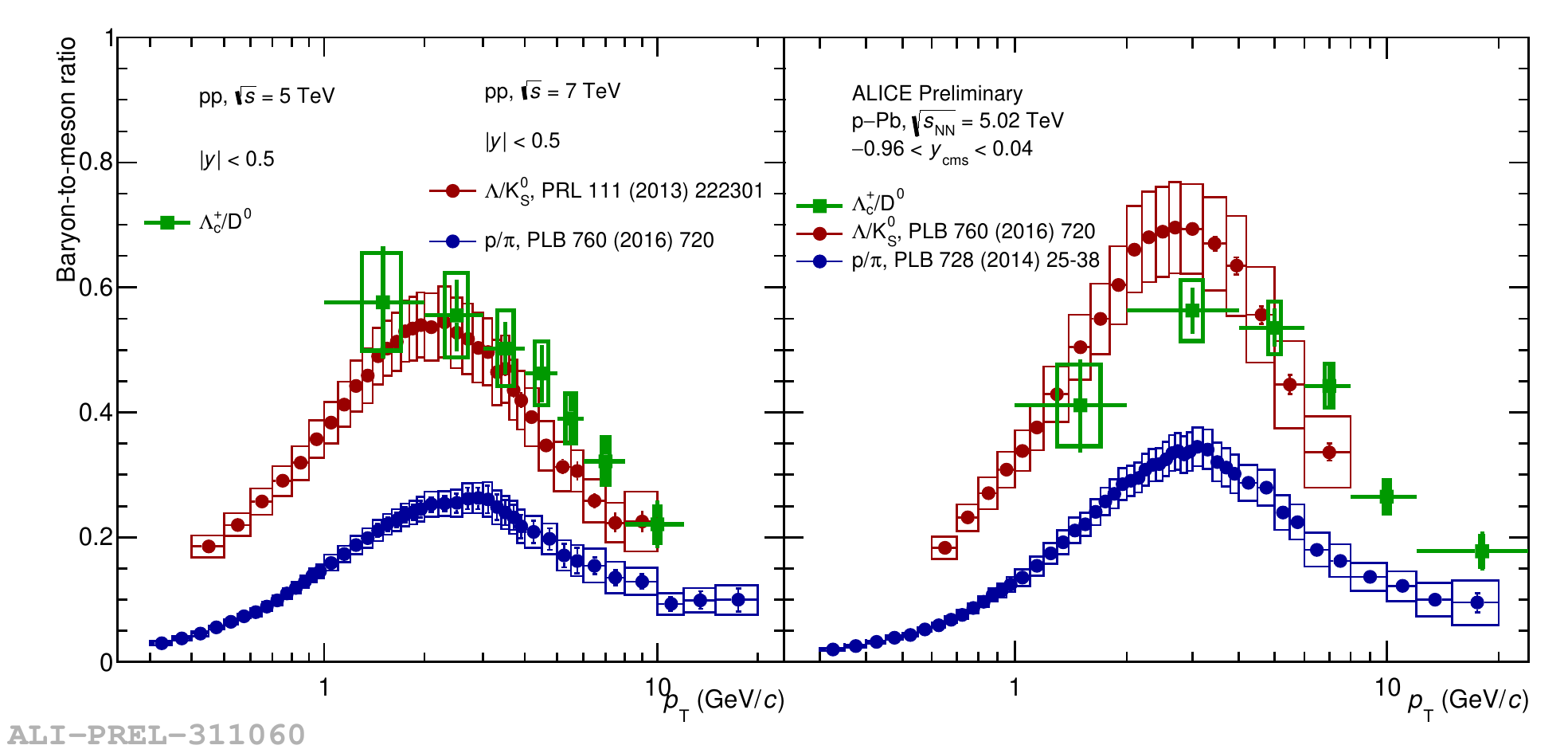}
	\caption{\lcp/\dzero ratios vs \pt\, in pp and p-Pb collisions compared to different event generators (upper panel) and to $\Lambda$/$\rm K^{0}_{S}$ and $\rm p$/$\pi$ ratios measured (lower panels).}
	\label{fig:lcVSd0}
	\end{center}
\end{figure}

In Fig.~\ref{fig:lcVSd0} (top panel) the \lc/\dzero ratio in pp collisions at $\sqrt{s}$ = 5 TeV and 7 TeV~\cite{Acharya:2017kfy} and in p-Pb collisions at $\sqrt{\rm s_{NN}}$ = 5 TeV is shown. The pp results are compared with predictions from Monte Carlo event generators, which are using fragmentations functions derived from measurements in $e^{+}e^{-}$ collisions. All the models underestimate the experimental results. Only PYTHIA 8 CR tune is able to reproduce the decreasing trend observed in data. ALICE results for the \lc/\dzero ratio are higher than previous measurements performed in $e^{+}e^{-}$ and $e\rm p$ collisions at lower centre-of-mass energies, in different rapidity and \pt \,intervals, where other processes may be involved. 
Moreover, differences observed in the beauty sector in p$\bar{\rm p}$ collisions at Tevatron~\cite{Aaltonen:2008zd} and LHC~\cite{Aaij:2015fea} with respect to $e^{+}e^{-}$ collisions at LEP~\cite{Gladilin:2014tba}, suggest a non universality of the fragmentation fractions. A decreasing trend for \lc/\dzero ratio is observed starting from \pt \,= 4 GeV/$c$ in both pp and p-Pb collisions. As shown in the bottom panels of Fig.~\ref{fig:lcVSd0}, the trend is similar to the baryon/meson ratio observed in the light-flavour sector. This could be a hint of independence of the baryon/meson ratio on the quark content, although definite conclusions can be drawn after improving the experimental uncertainties in the low-\pt \,region.
  
\begin{figure}[H]
	\begin{center}
		\includegraphics[width=0.9\linewidth]{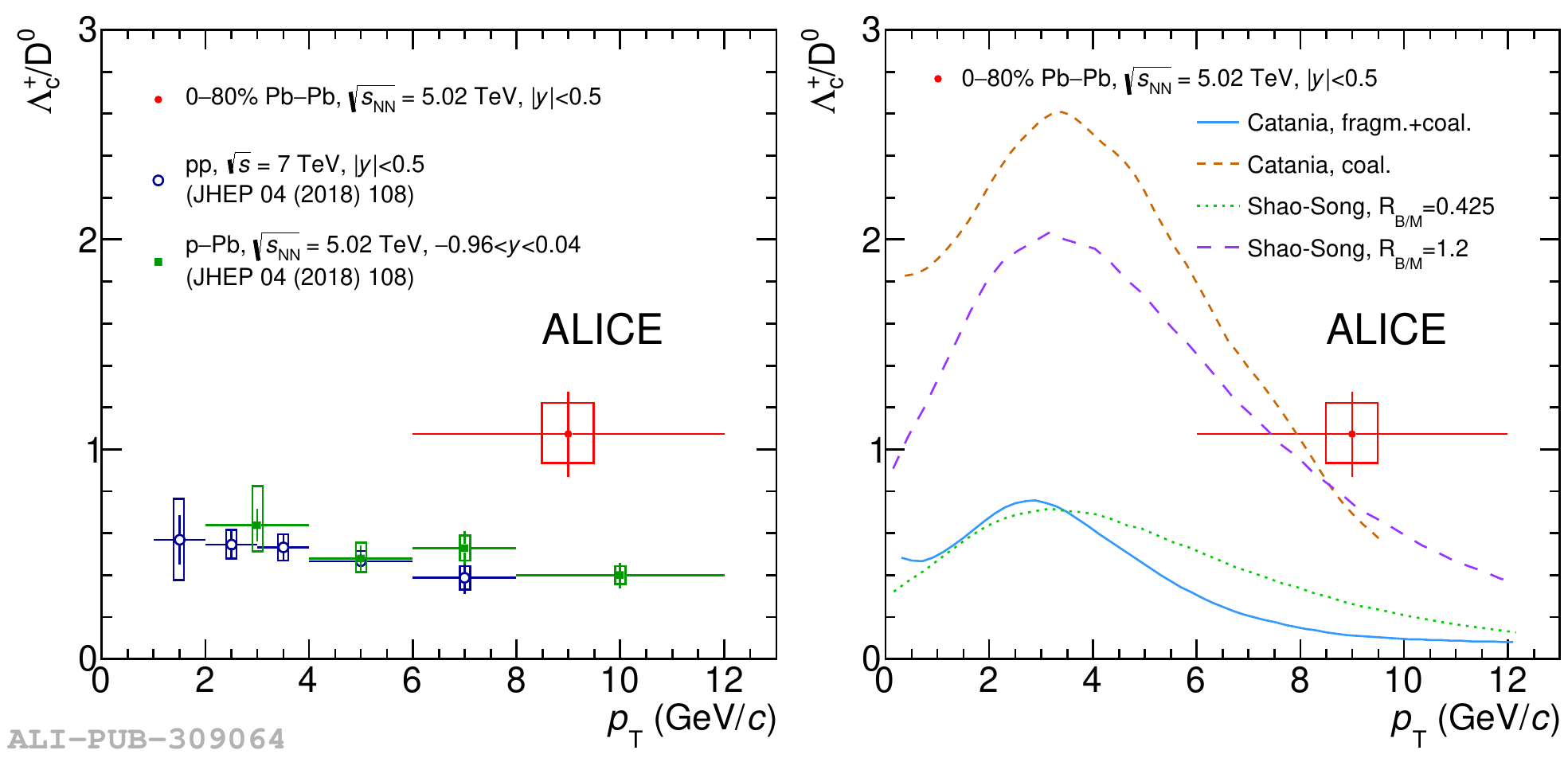}
		\caption{\lc/\dzero ratio as a function of \pt\, in 0-80\% most central Pb-Pb collisions compared with the measurements in pp and p-Pb collisions (left), and model calculations (right).}
		\label{fig:pbpb1}
	\end{center}
\end{figure}

The first measurement of prompt \lc production in Pb-Pb collisions at $\sqrt{\rm s_{NN}}$ = 5 TeV was performed recently by ALICE~\cite{Acharya:2018ckj}. The analysis was based on topological cut selection, using the hadronic decay channel \lcp$\rightarrow$$\rm pK^{0}_{s}$, in 0-80\% centrality class and in 6 $<$ \pt \,$<$$12$ GeV/$c$. The corresponding \lc/\dzero ratio, shown in Fig.\ref{fig:pbpb1}, is higher than that obtained in pp and p-Pb collisions. It is described by model calculations including only coalescence as \lc hadronisation mechanism~\cite{Plumari:2017ntm}.
 \section{Outlook} 
 The analysis of the Pb-Pb data collected at the end of 2018 will allow us to extend the \pt \,range and to have more differential measurements in \pt \,and centrality. Moreover, the ALICE upgrade (LHC Run 3 and 4), including an important upgrade of the ITS~\cite{Reidt:2014oma}, will allow for a better suppression of the background, increasing the precision of the measurements.

\end{document}